\documentclass[]{emulateapj}

\usepackage{graphicx}
\newcommand{\be}{\begin{equation}}
\newcommand{\ee}{\end{equation}}
\newcommand{\bea}{\begin{eqnarray}}
\newcommand{\eea}{\end{eqnarray}}

\begin{document}
\title{Non-locality of Hydrodynamic and Magnetohydrodynamic Turbulence}
\shorttitle{Locality of HD and MHD turbulence}
\shortauthors{Cho}

\author{Jungyeon Cho
}
\altaffiltext{1}{Dept. of Astronomy and Space Science, 
    Chungnam National Univ., Daejeon, Korea; cho@canopus.cnu.ac.kr}
%\altaffiltext{2}{Dept. of Astronomy, Univ. of Wisconsin, Madison, 
%    WI53706, USA; cho@astro.wisc.edu}

\begin{abstract}
We compare non-locality of interactions between different scales
in hydrodynamic (HD) turbulence and magnetohydrodynamic (MHD) turbulence
in a strongly magnetized medium.
We use 3-dimensional incompressible direct numerical simulations to evaluate non-locality of
interactions.
Our results show that non-locality in MHD turbulence
is much more pronounced than that in HD turbulence.
Roughly speaking, non-local interactions count for more than 10\% of total interactions 
in our MHD simulation on a grid of $512^3$ points. 
However, there is no evidence that non-local interactions are important
in our HD simulation with the same numerical resolution.
We briefly discuss how non-locality affects energy spectrum.
\end{abstract}

\section{Introduction}
Turbulence is commonly observed in astrophysical fluids and in many cases
such turbulence is accompanied by
a strong magnetic field, which has a large
impact on the dynamics of the turbulent cascade.
Since turbulence influences many astrophysical processes
(e.g.~transport of mass and angular momentum,
star formation, fragmentation of molecular
clouds, heat and cosmic ray transport, magnetic reconnection, etc.),
understanding scaling properties of magnetohydrodynamic (MHD) turbulence is
essential for theoretical astrophysics.
For this reason, rich literature exists regarding scaling relations of MHD turbulence
(See Goldreich \& Sridhar 1995; 
 Biskamp 2003 and references therein; see also Cho \& Vishniac 2000b; Maron \& Goldreich 2001;
 M\"uller, Biskamp, \& Grappin 2003;
M\"uller \& Grappin 2005;
Boldyrev 2005, 2006; Beresnyak \& Lazarian 2006; Mason, Cattaneo \& Boldyrev 2006;
 Gogoberidze 2007; Matthaeus et al. 2008).

In hydrodynamic (HD) turbulence, energy cascades down to smaller scales.
Kinetic energy contained in an ``eddy'' is transferred to smaller eddies
by shearing motions of other eddies (see, for example, Frisch 1995).
Most theories on turbulence assume locality of interactions, 
%%%which means that an eddy transfers its energy to smaller scale eddies
%%%via interactions with other similar size eddies.
which means
interactions between similar size eddies dominate in such energy cascade.
In Fourier space, this means that a Fourier mode at a wavenumber $k=|{\bf k}|$, where
${\bf k}$ is the wavevector, interacts mainly with other modes having similar wavenumbers and transfers
its energy to modes that have larger wavenumbers.
Recently many researchers have investigated locality in HD turbulence 
(Mininni, Alexakis, \& Pouquet 2008; Alexakis, Mininni, \& Pouquet 2007; see also Verma et al. 2005).

In MHD turbulence with a strong mean field (${\bf B_0}$), locality is also generally assumed.
However, in the MHD case, the nature of energy cascade is slightly different. %%% from its HD counterpart.
%%%For simplicity, let us suppose that
%%%a uniform external magnetic field (${\bf B}_0$) is present.
In the incompressible limit, any magnetic perturbation propagates
{\it along} the magnetic field line.
To the first order, the speed of propagation is constant and equal to
the Alfv\'en speed $V_A=B_0/\sqrt{4\pi \rho}$, where $\rho$ 
is the density.
Since wave packets are moving along the magnetic field line,
there are two possible directions for propagation. 
If all the wave packets are moving in one direction,
then they are stable to nonlinear order (Parker 1979).
Therefore, in order to initiate turbulence, there must be
opposite-traveling wave packets and the energy cascade occurs only when
they collide.
Therefore, in the MHD case, locality means that a wave packet (or ``eddy'') transfers energy to
smaller scale wave packets by shearing motions of opposite-traveling wave packets of similar size.

There have been some discussions about non-locality in MHD turbulence with a strong mean field.\footnote{
   When the mean field is weak or zero, turbulence structure is very different
   (see for example Cho et al. 2009).
   There are many discussions about non-locality in this regime %%%kind of turbulence
   (see for example Alexakis, Mininni, \& Pouquet 2005b; Lessinnes, Verma, \& Carati 2008; Yousef et al. 2009;
   Aluie \& Eyink 2010).
}
For example, Alexakis (2007) theoretically studied non-local model of MHD turbulence.
In their inspiring work, Beresnyak \& Lazarian (2010) numerically studied non-locality and 
                                         argued that ``MHD turbulence is
fairly non-local, at least less local than hydrodynamic turbulence'' (see also Beresnyak \& Lazarian 2009).
They claimed that `` a) the lack of
visible bottleneck effect in MHD turbulence, while it is
clearly present in hydro turbulence, and b) the dependence of kinetic and magnetic spectra on driving''
support this idea.
Teaca et al. (2009) calculated anisotropic energy transfer in Fourier space. But they did not 
pay much attention to the locality issue.

In this paper, we quantitatively evaluate non-locality of HD and MHD turbulence
and present a direct evidence that non-locality is
clearly present in MHD turbulence.
We consider only balanced strong MHD turbulence.
Here balanced MHD turbulence means that amplitudes of two opposite-traveling wave packets
are almost equal.
In \S2, we describe our numerical setup.
In \S3, we present our results.
In \S4, we briefly discuss how non-locality affects energy spectrum  and give summary.

\section{Simulations}
We solve the incompressible HD equation, %%%(Eq.~[\ref{veq}] with $B=0$)
\begin{equation}
\partial_t {\bf v}  = -(\nabla \times {\bf v}) \times {\bf v}
         + \nu \nabla^{2} {\bf v} + {\bf f} - \nabla P' ,
        \label{eq_hd}
\end{equation}
and the incompressible MHD equations, %%% in a periodic box of size $2\pi$:
\begin{equation}
\partial_t {\bf v}  = -(\nabla \times {\bf v}) \times {\bf v}
      + (\nabla \times {\bf B})
        \times {\bf B} + \nu \nabla^{2} {\bf v} + {\bf f} - \nabla P' ,
        \label{veq}
\end{equation}
\begin{equation}
\partial_t {\bf B} = {\bf B} \cdot \nabla {\bf v}
     - {\bf v} \cdot \nabla {\bf B} + \eta \nabla^{2} {\bf B},
     \label{beq}
\end{equation}
in a periodic box of size $2\pi$,
where $\bf{f}$ is a random forcing term with unit correlation time,
$P'\equiv P + v^2/2$, $P$ is pressure,
${\bf v}$ is the velocity, and ${\bf B}$ is the magnetic field divided by
$(4\pi \rho)^{1/2}$. Thus the field ${\bf B}$ is, in fact,
the Alfv\'{e}nic velocity. The velocity and the magnetic fields are 
divergence-free: 
$
      \nabla \cdot {\bf v} =\nabla \cdot {\bf B}= 0.
$
The peak of energy injection occurs at $k_L \approx 2.5$, 
so the energy injection scale is $L \sim 2.5$.
The amplitudes of the forcing components are tuned to ensure $v \approx 1$.

In the MHD simulation, the magnetic field consists of the uniform background field and a
fluctuating field: ${\bf B}= {\bf B}_0 + {\bf b}$.
The Alfv\'en velocity of the uniform background field, $B_0$, is set to 0.8.
At $t=0$, the magnetic field has only the uniform component.
We consider only the case where viscosity is equal to magnetic diffusivity: $
  \nu = \eta$.
Details of the code can be found in Cho \& Vishniac (2000ab). %%% or Cho \& Ryu (2009).

\begin{figure}
\centerline{
\includegraphics[width=0.45\textwidth]{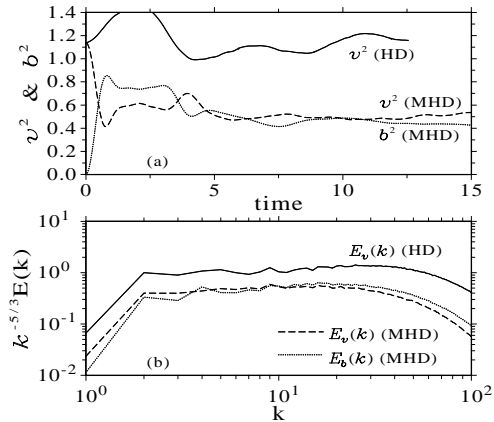}
}
\caption{(a) Time evolution of $v^2$ and $b^2$. 
         (b) Compensated spectra at $t=12$.}
\label{fig_t}
\end{figure}

%%%Upper panel of
Figure~\ref{fig_t}(a) shows time evolution of kinetic and magnetic energy densities.
%%%Lower panel of 
Figure~\ref{fig_t}(b) shows energy spectra at t=12.
The kinetic spectrum for HD Run (solid curve) is consistent with the Kolmogorov 
spectrum ($E(k)\propto k^{-5/3}$) for $k \in (2,15)$. But it shows a moderate increase of the slope for $k>15$.
The kinetic spectrum for MHD Run (dashed curve) is also consistent with the Kolmogorov one.
However, the magnetic spectrum (dotted curve) is slightly shallower than the Kolmogorov one.
Therefore, the spectrum of $b^2+v^2$ (not shown) is slightly shallower than the Kolmogorov one.

%%%%%%%%%%%%%%%%%%%%%%%%%%%%%%%%%%%%%%%%%%%%%% TABLE
%\clearpage
\begin{deluxetable}{llll} 
%\footnotesize
\tablecaption{Runs}
\tablewidth{0pt}
\tablehead{
   \colhead{Run} & 
   \colhead{Resolution} & 
   \colhead{$B_0$} & 
   \colhead{$\nu~(=\eta)$} 
}
\startdata 
HD        &  $512^3$  &  -    &  .0004 \\
MHD       &  $512^3$  &  0.8  &  .0004 
%             \hline
\enddata
\label{table_1} 
\end{deluxetable}
%%%%%%%%%%%%%%%%%%%%%%%%%%%%%%%%%%%%%%%%%%%%%% TABLE

%%%%%%%%%%%%%%%%%%%%%%%%%%%%%%%%%%%%%% fig
%%%%%%%%%%%%%%%%%%%%%%%%%%%%%%%%%%%%%% fig

\section{Results}
\subsection{Shell-to-Shell Energy Transfer}

%%%%%%%%%%%%%%%%%%%%%%%%%%%%%%%%%%%%%% fig
\begin{figure*}
\centerline{
\includegraphics[angle=-90,width=0.95\textwidth]{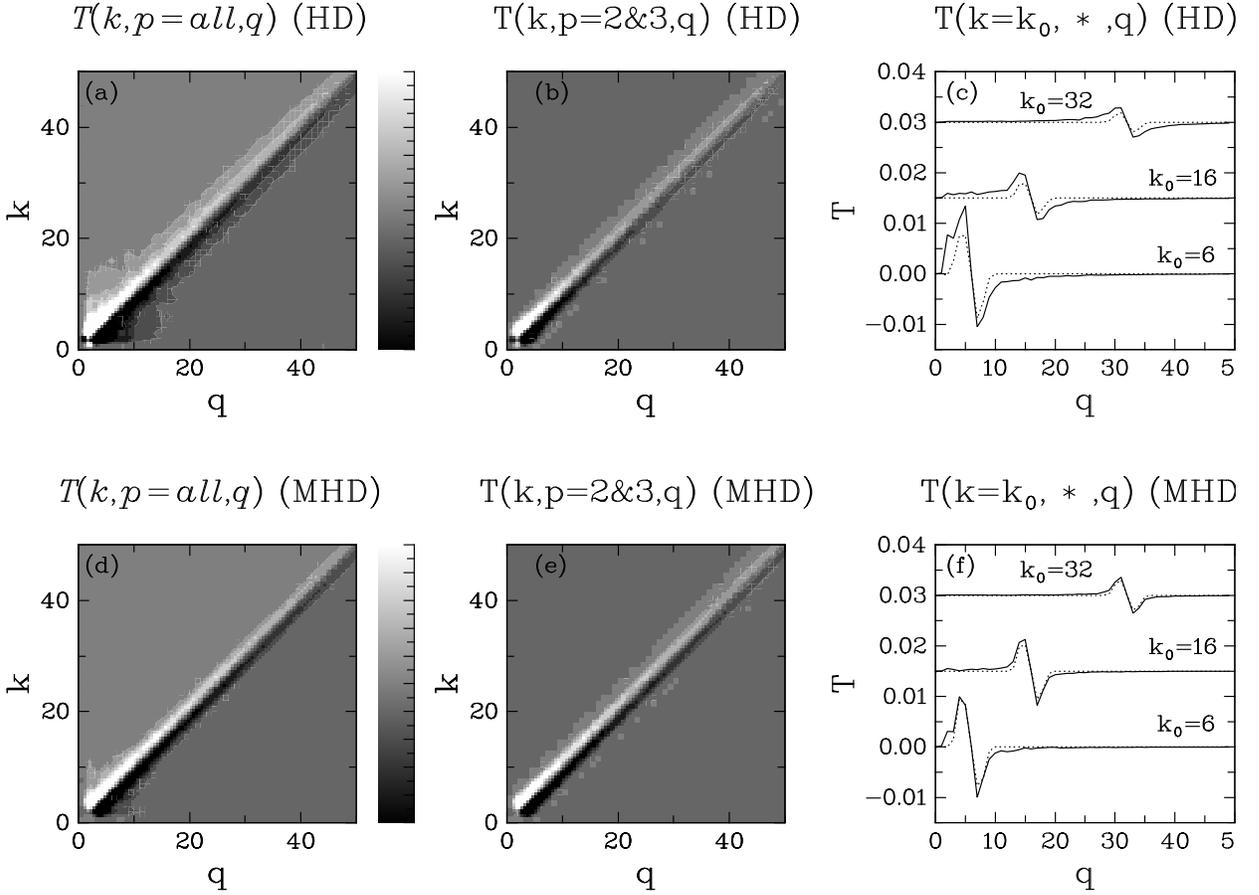}
}
\caption{(a) \& (d): Energy transfer from $q$-shell to $k$-shell by mediation of all $p$ modes
  that satisfy ${\bf p}={\bf k}-{\bf q}$.
  (b) \& (e): Energy transfer from $q$-shell to $k$-shell by mediation of $p$ modes
  that satisfy $1.5 \leq |{\bf p}| < 3.5$. This is equal to the Energy transfer from $q$-shell to $k$-shell
  mediated by the outer scale motions.
  (c) \& (f): $T(k_0,*,q)$ vs. $q$ for $k_0=6, 16,$ and $32$.
              The values for $k_0= 16$ and $32$ are offset by 0.015 and 0.03, respectively.
              The solid lines are for $T(k_0,p=\mbox{all},q)$ and the dotted lines
              for $T(k_0,p=2\&3,q)$.
              In the MHD case, the solid and the dotted lines look very similar,
              which might mean that the outer scale counts for substantial fraction of
              the energy transfer rate.
  In contour diagrams the linear color bars range from -0.005 to 0.005.
  Snapshot at $t=12$. 
  }
\label{fig_qk}
\end{figure*}
%%%%%%%%%%%%%%%%%%%%%%%%%%%%%%%%%%%%%% fig

We can rewrite the MHD equations in Eqs.~(\ref{veq}) and (\ref{beq}) using the 
Els\"asser variables, ${\bf Z}^+\equiv {\bf v}+{\bf B}$ and ${\bf Z}^-\equiv {\bf v}-{\bf B}$:
\bea
\partial_t {\bf Z}^+  =  -{\bf Z}^- \cdot  \nabla {\bf Z}^+
       +\nu \nabla^{2} {\bf Z}^+ + {\bf f} - \nabla P ,
        \label{eq_zplus}  \\
\partial_t {\bf Z}^-  =  -{\bf Z}^+ \cdot  \nabla {\bf Z}^-
       +\nu \nabla^{2} {\bf Z}^- + {\bf f} - \nabla P.
        \label{eq_zminus}          
\eea
The Els\"asser variables denote amplitudes of two opposite-traveling waves along
the magnetic field line.
The nonlinear term of ${\bf Z}^+$, for example, states that energy transfer
between ${\bf Z}^+$ modes is mediated by ${\bf Z}^-$ modes.

In Fourier space, the non-linear term in Eq.~(\ref{eq_zplus}), for example, becomes
\be
       {\bf N}_{\bf k}^+ \equiv
       -i{\bf k} \cdot \sum_{ {\bf p}+{\bf q}={\bf k}} {\bf Z}_{\bf p}^- {\bf Z}_{\bf q}^+
       \label{eq_6}
\ee
and the time derivative of $(1/2)|{\bf Z}_{\bf k}^+|^2$ is given by
\be
    {\bf Z}_{\bf k}^+ \cdot \partial_t {\bf Z}_{\bf k}^{+*} 
    =  {\bf Z}_{\bf k}^+ \cdot {\bf N}_{\bf k}^{+*}
       -\nu k^{2} |{\bf Z}_{\bf k}^+ |^2,
        \label{eq_7} 
\ee
where `*' denotes complex conjugate and
we dropped the forcing term because its role is limited in the inertial range.
%%%The ${\bf z}_{\bf p}^-$ modes do not gain or lose energy.
Energy transfer occurs only between ${\bf Z}_{\bf k}^+$ and ${\bf Z}_{\bf q}^+$, via
shearing motions provided by ${\bf Z}_{\bf p}^-$ modes.
Without ${\bf Z}^{-}$ modes, ${\bf Z}^{+}$ modes alone do not interact each other.

If interactions are local in Fourier space, we will have $p\sim q\sim k$.
Since it is difficult to check the locality using individual triad interactions in Fourier space, 
we investigate shell-to-shell interactions.
That is, we consider collective interactions in Fourier space between ${\bf Z}^+$ modes in a  
unit shell of
radius $k$ (hereinafter, ``$k$-shell'') 
and ${\bf Z}^{+}$ modes in a unit shell of radius $q$ (``$q$-shell'')
by the help of ${\bf Z}^{-}$ modes in a unit shell of radius $p$ (``$p$-shell'').\footnote{
    In this paper, wavenumbers $p$, $q$, and $k$ refer to those described here (and those appear in 
        Eqs.~(\ref{eq_6}), (\ref{eq_7}) and (\ref{eq:Tdef})).
        }

We first consider shell-to-shell interactions of ${\bf Z}^{+}$ modes mediated 
by {\it all} ${\bf Z}^{-}$ modes.
Contour diagrams in Figure~\ref{fig_qk}(a) and (d) show the shell-to-shell energy transfer rate:
\begin{eqnarray}
     && \mathcal{T}(k,p=all,q) \equiv \sum_{p=0}^{p_{max}} \mathcal{T}(k,p,q) =  \label{eq:Tdef} \\
     && \sum_{k-1/2<|{\bf k}^\prime|<k+1/2} ~~\sum_{q-1/2<|{\bf q}^\prime|<q+1/2}   
        {\bf Z}^{+}_{{\bf k}^{\prime}} \cdot \left[ 
        i {\bf k}^\prime \cdot %%%\sum_{ {\bf p}+{\bf q}^\prime={\bf k}^\prime } 
        \left( {\bf Z}_{\bf p}^- {\bf Z}_{{\bf q}^\prime}^+  \right)^* \right],
        \nonumber
\end{eqnarray}
where ${\bf p}={\bf k}^\prime -{\bf q}^\prime$ and $p_{max}$ is the largest wavenumber,
for MHD (Figure~\ref{fig_qk}(d)) and a similar expression for HD (Figure~\ref{fig_qk}(a)).\footnote{
        In the HD case, ${\bf Z}^+={\bf Z}^-={\bf v}$.
        }
The energy transfer rate $\mathcal{T}(k,p=all,q)$ here is similar to $T_2(K,Q)$ 
in Alexakis, Mininni, \& Pouquet (2005a)
or Mininni et al. (2008).
The contour diagrams are exactly anti-symmetric with respect to the $k=q$ line.
The overall shape of the contour diagram for HD is consistent with earlier
findings (Alexakis et al. 2005a; Mininni et al. 2008).
The value of $\mathcal{T}(k,p=all,q)$ is positive on the upper-left half, 
which means that, when $q<k$, the ${\bf Z}^{+}$ modes in $k$-shell gains energy from 
the ${\bf Z}^{+}$ modes in $q$-shell by the help of all 
${\bf Z}^{-}$ modes that satisfy ${\bf p}+{\bf q}={\bf k}$.
This result is consistent with the concept of energy cascade: energy cascades down to smaller scales.
Note that the values of $\mathcal{T}(k,p=all,q)$ are very close to zero except near the $k=q$ line.
Does it mean that locality is a good approximation?

%%%%%%%%%%%%%%%%%%%%%%%%%%%%%%%%%%%%%% fig
\begin{figure*}
\centerline{
\includegraphics[angle=-90,width=0.90\textwidth]{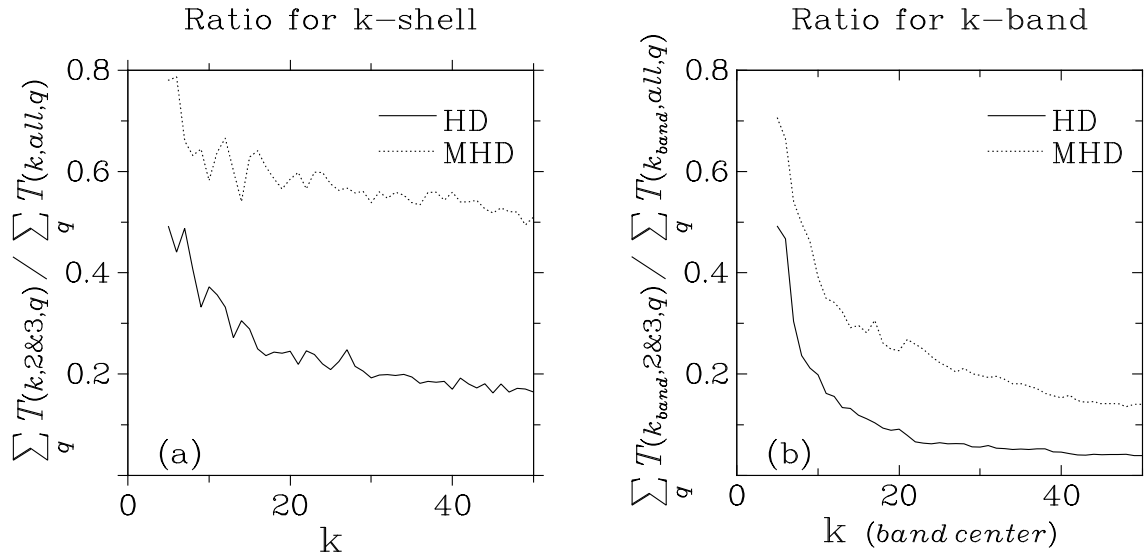}
}
\caption{  Ratio of local to non-local interactions.   %%% energy transfer rates.
    (a) The ratio of energy transfer from all $q$-shells with $q<k$ to $k$-shell
  mediated by the outer scale motions ($p=2$ and $3$) to that mediated by all scale motions (all $p$-shells).
    (b) The ratio of energy transfer from all $q$-shells with $q<k_{min}=k/\sqrt{2}$ to all $k$-shells
    between $k/\sqrt{2}$ and $\sqrt{2}k$ (we call ``$k$-band''),
  mediated by the outer scale motions ($p=2$ and $3$) to that mediated by all scale motions (all $p$-shells).
  Snapshot at $t=12$. }
\label{fig_ratio}
\end{figure*}
%%%%%%%%%%%%%%%%%%%%%%%%%%%%%%%%%%%%%% fig

Note that, when the outer scale of turbulence
provides strong shearing motions, $\mathcal{T}(k,p=all,q)$ has positive peaks at $q=k-k_L$ and
negative peaks at $q=k+k_L$. In our case, $k_L \sim 2.5$.
Therefore, shearing motions of the outer scale can also produce diagrams similar to Figure~\ref{fig_qk}(a) and (d).
Indeed, when we plot 
\be
  \mathcal{T}(k,p=2 \& 3,q)\equiv \sum_{p=2}^3 \mathcal{T}(k,p,q), 
\ee
which is similar to  $\mathcal{T}(k,p=all,q)$ except 
the fact that we do the summation from $p=2$ to $p=3$, 
%%%the fact that we do the summation (see Eq.~[\ref{eq:Tdef}]) for $1.5 \leq p <3.5$,
the contour diagrams show similar features (Figure~\ref{fig_qk}(b) and (e)).
This result is consistent with earlier results for HD turbulence (e.g.~Alexakis et al.~2005a).
Note that, in Figure~\ref{fig_qk}(a) and (d), the width of the contour lines near the $k=q$ line 
is narrower in the MHD case than in the HD case.
This might mean that the effect of the outer scale is stronger in the MHD case than in the HD case.

Figure~\ref{fig_qk}(c) and (f) (right panels) show the values of 
$T(k_0,p=all,q)$ (solid) and $T(k_0,p=2\&3,q)$ (dotted) for select values of $k_0$.
We take $k_0=6, 16,$ and $32$.
The values for $k_0= 16$ and $32$ are offset by 0.015 and 0.03, respectively, for clarity.
In the HD case (upper panel), 
$T(k_0,p=all,q)$ and $T(k_0,p=2\&3,q)$ look different.
However, in the MHD case (lower panel),
$T(k_0,p=all,q)$ and $T(k_0,p=2\&3,q)$ look very similar, which might mean
that the outer scale {\it does} play important roles in 
              MHD energy cascade.

In order to evaluate the role of the outer scale in shell-to-shell energy transfer,
we calculate the ratio
\be
     \sum_{q=0}^{k-1} \mathcal{T}(k,p=2 \& 3,q) / \sum_{q=0}^{k-1} \mathcal{T}(k,p=all,q),
   %%%  \sum_{q=k}^{q_{max}} \mathcal{T}(k,p=2 \& 3,q) / \sum_{q=k}^{q_{max}} \mathcal{T}(k,p=all,q) \mbox{~(dotted)}
     \label{eq:shell_ratio}
\ee
where $q_{max}$ is the largest wavenumber.
%%%When the ratios are larger than 0.5,
%%%non-local interactions dominate.
Figure~\ref{fig_ratio}(a) shows the ratios for HD and MHD.
In the HD case (solid curve), the ratio is less than 0.5 for most values of $k$,
which is consistent with Mininni et al.~(2008).
However, in the MHD case (dotted curve) the values are $\gtrsim 0.5$
for most values of $k$, which means that 
non-local interactions are indeed important for shell-to-shell energy transfer in the MHD case.

However, it is very important to note that the result in Figure~\ref{fig_ratio}(a) does {\it not}
mean that non-local interactions are as strong as local interactions in MHD cascade.
The result in Figure~\ref{fig_ratio}(a) is only for a single shell.

In order to evaluate non-locality,
we would better consider the effect of $p$-shells on a band of $k$-shells between $k_{min}=k/\alpha$ 
and $k_{max}=\alpha k$, where
%%%$k_{min}=$, $k_{max}=$, and
$\alpha$ is a constant. 
In this paper, we take $\alpha=\sqrt{2}$.
The motivation for considering this quantity is that Fourier modes in $(k/\sqrt{2},\sqrt{2}k)$
can define ``eddies'' on a scale $l\sim 1/k$.
In Figure~\ref{fig_ratio}(b) we plot the ratio similar to that in Eq.~(\ref{eq:shell_ratio}), but expressed
in terms of  
\be
 \mathcal{T}(k_{band},\ldots,\ldots)\equiv \sum_{k^\prime=k_{min}}^{k_{max}} \mathcal{T}(k^\prime,\ldots,\ldots),
\ee
where $k_{min}=k/\sqrt{2}$ and $k_{max}=\sqrt{2}k$. The summation for $q$ is done from $0$ to $k_{min}-1$.
The ratio for MHD (dotted) is non-negligible and still substantially larger than that for HD (solid).
Therefore, we can conclude that non-locality is indeed present in MHD turbulence.\footnote{
   We note that the ratio for MHD gradually decreases as $k$ increases.
   Although it is not very clear at this moment whether
   it will continue to drop when we have a very long inertial range,
   it is likely
   that the ratio will continue to drop and the non-local effects of the outer scale will ultimately vanish on very small scales.
   Nevertheless, non-locality is an important characteristic of MHD turbulence near the outer scale.
}

\subsection{More on Non-locality of MHD Turbulence}
\begin{figure}
\centerline{
\includegraphics[angle=-90,width=0.45\textwidth]{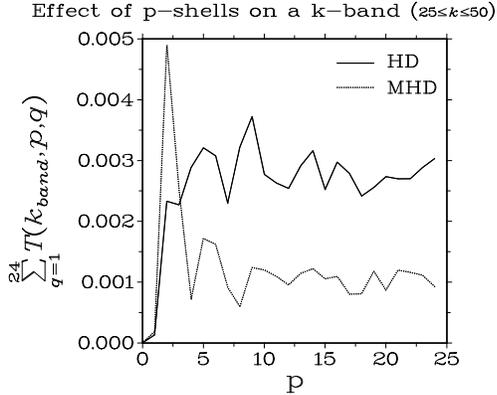}
}
\caption{  
    The amount of energy transferred to a $k$-band ($25 \leq k \leq 50$)
    from all $q$-shells having wavenumbers smaller than 25 (i.e. $q \leq 24$)
    as a function of $p$, which is the radius of a $p$-shell in Fourier space.
    Solid line is for HD turbulence and dotted line for the MHD case.
    In HD, non-locality is not conspicuous.
    In MHD, we can see a sharp peak at $p=2$.
    Snapshot at $t=12$.
    }
\label{fig_pq}
\end{figure}

Figure~\ref{fig_ratio}(b) shows that the $p=2$ and $p=3$ shells contribute
more than 10\% of the total energy flux. %%% to any $k$-band.
Then, which $p$-shell provides the strongest contribution to a $k$-band?
In other words, what is the most shear-providing shell for a band of $k$-shells between $k/\sqrt{2}$ and $\sqrt{2}k$?
To see this, we calculate the following quantity:
\be
  \sum_{q=0}^{24} \mathcal{T}(k_{band},p,q)\equiv \sum_{q=0}^{24} \sum_{k^\prime=25}^{50} \mathcal{T}(k^\prime,p,q),
\ee
which is equal to the total energy transferred from all $q$-shells with $q \leq 24$ 
to the $k$-band between $k=25$ and $k= 50$
by the shearing action of a $p$-shell.  %%% (Figure~\ref{fig_pq}).
Figure~\ref{fig_pq} shows that
each $p$-shell provides a similar contribution in the HD case (solid line).
Therefore, non-locality does not seem to be important in HD turbulence.
However, the $p=2$ shell contributes most in the MHD case (dotted line).
This is another piece of evidence that non-locality is clearly present in MHD turbulence.

%%%%%%%%%%%%%%%%%%%%%%%%%%%%%%%%%%%%%% fig
%%%%%%%%%%%%%%%%%%%%%%%

%%%%%%%%%%%%%%%%%%%%%%%
\begin{figure*}
\centerline{
\includegraphics[angle=-90,width=0.90\textwidth]{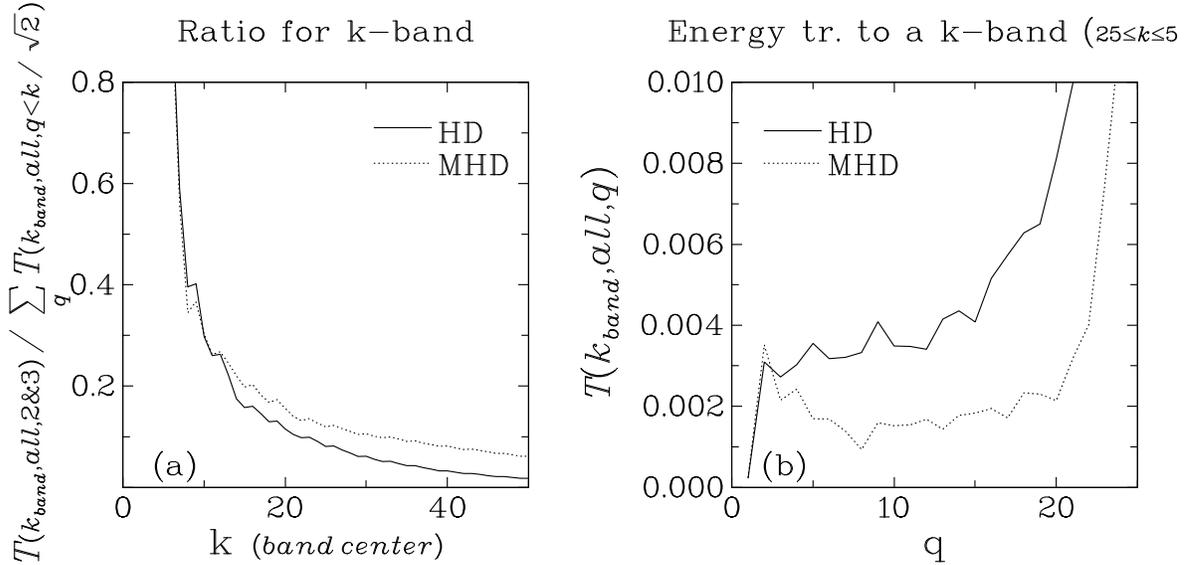}
}
\caption{Non-local energy transfer.
    (a) Ratios of local to non-local direct energy transfer rates.
    The quantity shown here, $T(k,p=all,q=2\&3)/T(k,all,0\leq q < k_{min})$,
    is the ratio of non-local energy transfer rate to total energy transfer rate.
    The numerator is the rate from $q=2$ and $q=3$ shells to $k$-bands between
     $k_{min}=k/\sqrt{2}$ and  $k_{max}=\sqrt{2}k$ and
     the denominator is the 
    total energy transfer rate from $q$-shells between $0$ and $k_{min}-1$ 
    to the same $k$-bands.
    The MHD case shows stronger non-locality.
    (b) The amount of energy transferred from a $q$-shell to a $k$-band ($25 \leq k \leq 50$)
    by the mediation of all $p$-shells.
    Solid line is for HD turbulence and dotted line for MHD turbulence.
    In both cases, non-locality is not conspicuous.
    Snapshot at $t=12$.
    }
\label{fig_energy}
%\caption{
%  The energy transfer spectra for the MHD case.
%  The solid curve ($T_{b\rightarrow v(k)}$) denotes 
%  energy gained by velocity components within a unit shell of
%  radius $k$ in wavevector space through interactions with the
%  magnetic field. The dotted curve ($T_{v\rightarrow b(k)}$) is
%  similarly defined.
%  The curves imply that the outer scale motions excite
%  small-scale velocity and small-scale magnetic field.
%  Snapshot at $t=12$.
%\label{fig_tvb}
%}
\end{figure*}
%%%%%%%%%%%%%%%%%%%%%%%%%%%%%%%%%%%%%% fig

\subsection{Non-local energy transfer}   \label{sect:et}
So far, we have discussed non-local influence of the outer scale eddies.
Now, it is time to clarify the meaning of non-locality.
The nonlinear term for $\partial_t {\bf Z}^+$, for example,
is $ -{\bf Z}^- \cdot  \nabla {\bf Z}^+$.
Since the nonlinear term contains both ${\bf Z}^-$ and ${\bf Z}^+$, non-locality has two meanings:
\begin{enumerate}
\item Non-local effects of ${\bf Z}^-$, and
\item Non-local effects of ${\bf Z}^+$.
\end{enumerate}
Since ${\bf Z}^-$ modes do not lose or gain energy, the former type of non-locality
does not involve energy transfer between the outer scale and small scales.
In fact, when this kind of non-locality is present,
energy transfer between adjacent shells is enhanced.  
Non-locality we have discussed so far is this type of non-locality.

When the latter type of non-locality is present, there {\it is}
direct energy transfer between different scales.
In order to evaluate the energy transfer rate from the outer scale 
to a $k$-band between $k_{min}=k/\sqrt{2}$ and $k_{max}=\sqrt{2}k$,
we calculate the ratio
\be
     \frac{ \mathcal{T}(k_{band},p=all,q=2\&3) }{ \sum_{q^\prime=0}^{k/\sqrt{2}-1} \mathcal{T}(k_{band},p=all,q^\prime)}.
     \label{eq:etratio}
\ee
Figure~\ref{fig_energy}(a) shows the ratios for HD and MHD.
We can see that the ratios for this type of non-locality are smaller than those for the former type of non-locality
(see Figure~\ref{fig_ratio}(b)).
The ratio for the MHD case is higher than that for the HD case.

Figure~\ref{fig_energy}(b) shows the values of
\be
    \mathcal{T}(k_{band},p=all, q)\equiv \sum_{k^\prime=25}^{50} \mathcal{T}(k^\prime,p=all,q)
\ee
 for HD and MHD.
The values of $\mathcal{T}(k_{band},p=all, q=2)$ and $\mathcal{T}(k_{band},p=all, q=3)$
are not particularly larger than other values.
It is clear from the figure that energy transfer from
the outer scale ($q=2$ and $3$) to the $k$-band ($25 \leq k \leq 50$)
is small. 
Therefore,  
non-local energy transfer from the outer scale to small scales
may not be an important characteristic for both the HD and the MHD cases.

\section{Discussion and Summary}
When shearing motions of the outer scale eddies influence energy transfer of
inertial range eddies, energy spectrum becomes flatter than the Kolmogorov one.
Suppose that the shearing motions of the outer scale eddies completely dominate energy cascade.
In this case, from 
\be
   Z_l^2/t_{cas} \sim Z_l^2/(L/v_L) \propto Z_l^2 = constant,
\ee
we can easily show that energy spectrum is $E(k)\propto k^{-1}$,
where $Z_l$ is an Els\"asser variable at scale $l$, $L$ the outer scale, and $v_L$ the rms velocity at the
outer scale (see Equation (1) of Cho, Lazarian, \& Vishniac 2003).
If the shearing motions of the outer scale eddies do not completely dominate,
we will have a spectrum between $k^{-1}$ and $k^{-5/3}$.

Indeed, in Figure~\ref{fig_t} we observe that $E_v(k)+E_b(k)$ 
(hence spectrum of ${\bf Z}^+$ or ${\bf Z}^-$) in MHD
is flatter than the Kolmogorov spectrum.
This is consistent with earlier numerical results 
(see, for example, Maron \& Goldreich 2001; M\"uller et al. 2003).

In summary, we have found the following results.
\begin{enumerate}
\item We have developed a quantitative method to measure non-locality (see
      Figure~\ref{fig_ratio}(b) or Figure~\ref{fig_pq}).

\item Our numerical calculations show 
     non-locality is more pronounced in MHD turbulence than in HD turbulence.
     This result confirms an earlier finding by Beresnyak \& Lazarian (2010).
     
\item There are two forms of non-locality in MHD turbulence: non-local influence of
      shearing motions, which does {\it not} involve energy transfer between different scales,
       and non-local energy transfer between different scales (\S\ref{sect:et}).
       In MHD, the former type of non-locality (i.e. non-local influence of outer scale shearing motions)
       is more important.
       It is not clear whether the latter type is important.
%     Therefore, non-local interactions in MHD means non-local 
%     influence of the outer scale shearing motions, 
%     rather than
%     non-local energy transfer.
     
\item In MHD, non-locality is not negligible so that it might affect
      dynamics of turbulent cascade.
   
\end{enumerate}
%%%As a final note we should emphasize that, in this paper, by non-local interactions we mean non-local 
%%%influence of the outer scale shearing motions, 
%%%rather than
%%%non-local energy transfer.
%%%However, we may need higher numerical resolution is needed.

\acknowledgements
J.C.'s work was supported by the National Research Foundation of Korea (NRF)
 grant
funded by the Korean Government (MEST) (NO. 2009-0077372). 
The works of J.C. was also supported by KICOS through
the grant K20702020016-07E0200-01610 provided by MOST.

\end{document}